\documentclass{agujournal}
\drafttrue{}
\usepackage{amsfonts,amsmath,amssymb,enumerate,color,mathtools,mathrsfs}
\usepackage{siunitx}
\usepackage{cleveref}
\usepackage{chemformula}
\usepackage{physics}
\usepackage{graphicx}
\usepackage{booktabs}
\usepackage{multirow}

\journalname{Geophysical Research Letters}

\begin{document}

\title{From the seep to the surface: the ascent and dissolution of methane bubbles in the ocean}

%
%




\authors{Jiangzhi Chen\affil{1}}

\affiliation{1}{Institute of Deep Sea Science and Engineering, Chinese Academy of Sciences, Sanya, Hainan Province 572000}

\correspondingauthor{Jiangzhi Chen}{chenjz@idsse.ac.cn}

\begin{keypoints}
\item a spherical bubble ascent model is developed to estimate the amount of methane gas in the oceans entering the atmosphere
\item the model is employed to understand the ocean-air methane exchange in the Shenhu area, South China Sea
\item model results are used to quantify the possible outcomes of methane gas leak caused by the dissociation of methane hydrate
\end{keypoints}

\begin{abstract}
 Methane, as a strong greenhouse gas, has 21--25 times the warming potential per unit mass than carbon dioxide, and the methane from the oceans can contribute to $\sim4\%$ of the annual atmosphere methane budget. Large methane bubble plumes have been observed in seep sites globally on shallow continental shelves, and emerging industry of methane hydrates mining causes growing environmental concern on possible disastrous blowout which destabilizes the methane hydrate and releases huge amount of methane gas. To better estimate how much methane in gaseous phase leaked from the seeps can reach the atmosphere, a simplified model is developed to simulate the ascent of a methane bubble from a shallow ocean methane seep, and the methane transfer with the surrounding water. The breakup and coalescence of bubbles are neglected, and the bubble is assumed to remain spherical following a vertical path during the whole rising process. We calculated the survival distance of bubbles with varying initial sizes and depths and the remaining percentage of methane reaching the sea surface, and applied the results to the seep sites in the Shenhu area in the South China Sea. The study can provide insight into the relative significance of different water bodies in contributing to the atmosphere greenhouse gas.
\end{abstract}

\section{Introduction}

Atmospheric methane (\ch{CH4}) is a strong greenhouse gas, with at least 20 times the warming potential per unit mass than that of \ch{CO2} \citep{mcginnis2006fate,leifer2010characteristics}. Observations have confirmed that the concentration of methane has tripled since preindustrial times \citep{bousquet2006contribution}, and is increasing fast in the atmosphere \citep{dlugokencky2011global,sussmann2012renewed,nisbet2016rising}. Currently about 64\% of methane released to the atmosphere comes from anthropogenic sources (e.g., livestock farming, fossil fuels, and biomass burning), but natural sources (e.g., wetlands, termites, lakes, and oceans) also contribute to about 36\% of total methane budget \citep[][see supplemental material]{bousquet2006contribution}, among which oceans comprise up to 10\% of the natural methane emission, and lakes provide another 10\% \citep{bastviken2004methane}.

In oceans, methane is produced by microbial activities mainly related to degradation of the organic material in the sediments and serpentinization and iddingsization where hydrogen is produced. Seep sites with large volumes of methane gas leakage have been reported globally via visual and acoustic methods. These sites vary in depth with equivalent methane bubble radii less than \SI{0.5}{\cm} (see \Cref{table:flux}). The small radius may be determined by the sizes of sediment particles and the methane flux rates.
\begin{table}[h]
 \centering
 \begin{tabular}{@{}lllll@{}}
  \toprule
  \multicolumn{1}{c}{Site} & Depth (\si{\m}) & $a_e$ (\si{\mm}) & Reference \\\midrule
  Central Nile Deep Sea Fan & $\sim$1650 & 2.75 & \citet{roemer2014methane} \\
  H\r{a}kon Mosby mud volcano (Norway) & 1250 & 2.6 & \citet{sauter2006methane} \\
  GC-185 in the Gulf of Mexico & \numrange{525}{550} & $\sim$3 & \citet{leifer2003dynamics} \\
  Makran continental margin (Pakistan) & \numrange{575}{2870} & 2.6 & \citet{roemer2012quantification} \\
  Vodyanitski mud volcano (Black Sea) & 2070 & 2.6 & \citet{sahling2009vodyanitskii} \\
  Kerch seep area (Black Sea) & 890 & 2.89 & \citet{roemer2012geological} \\
  northwestern Black Sea shelf & \numrange{70}{112} & 2.6 & \citet{greinert2010atmospheric} \\
  southern summit of Hydrate Ridge, Oregon & 780 & \numrange{3}{3.5} & \citet{rehder2002enhanced} \\
  Utsira High, Central North Sea (Norway) & \numrange{81}{93} & \numrange{1.6}{3.7} & \citet{vielstaedte2015quantification} \\ \bottomrule
 \end{tabular}
 \smallskip
 \caption{Published global seeps sites where methane bubble emission is observed. The depth and equivalent bubble radius are reported here. The depth of sites vary between \SIrange{70}{2870}{\m}, but the bubble radii are all below \SI{4}{\mm}.}\label{table:flux}
\end{table}
Besides the gaseous and dissolved methane, there is huge amount of methane ($\sim$\SI{1e4}{\giga\tonne}) stored in ocean sediments in the form of methane hydrate \citep{kvenvolden1988methane}. Recently the economic potential of methane hydrate as a possible energy source has attracted more attention, and several countries have conducted experiments to mine ocean methane hydrate, which causes growing environmental concern on possible disastrous blowout with massive dissociation of methane hydrate and huge methane gas emission, similar to the Deepwater Horizon disaster in the Gulf of Mexico or 22/4b in the North Sea. Besides impacts on global warming, huge methane emission can greatly reduce the ocean pH by means of anaerobic methane oxidation, and destroy the ocean environment \citep{dickens1995dissociation,hesselbo2000massive}. The positive feedback between global warming and methane hydrate dissociation will further deteriorate the situation \citep{thomas2002warming}.

To better quantify the contributions of ocean methane to global warming and estimate the risk of massive methane hydrate dissociation, it is important to understand the methane gas transport in oceans. The remaining percentage of methane in a bubble mainly depends on the initial depth and the size. We developed a simplified model simulating the ascent of a methane bubble from a shallow seep located on a continental shelf, e.g., in the Shenhua area in the South China Sea, and model the methane transfer with the surrounding seawater. The split and coalescence of bubbles are neglected, and bubbles are assumed to remain spherical following vertical paths during the whole ascent. We calculated the survival distance of bubbles with varying initial sizes and depths, and the percentage of methane remained when bubbles reach the ocean surface.

\section{Regional setting}

Shenhu area, near the Pearl River Mouth Basin in the northern continental margin of the South China Sea, has been recognized as a promising place of gas hydrate extraction due to its relatively high hydrate saturation. Particularly, samples collected from the Shenhu area contain high saturation hydrates (the fraction of pore spaces occupied by gas hydrates) up to \SI{45}{\percent} \citep{wang2011elevated}, drastically higher than that of some other ocean gas hydrate sites where the saturations are only \SIrange{1}{8}{\percent} \citep{davie2001numerical,davie2003sources}. These samples were discovered in fine-grained sediments about \SI{1200}{mbsf}. In May, 2017, China successfully extracted gas from the hydrates in a series of production runs in this region, and it is reasonable to believe that more further field explorations will be conducted here, so we apply our model to the Shenhu area to calculate the survival time and distance of methane bubbles.

\begin{figure}[h]
 \centering
 \includegraphics[width=0.5\textwidth]{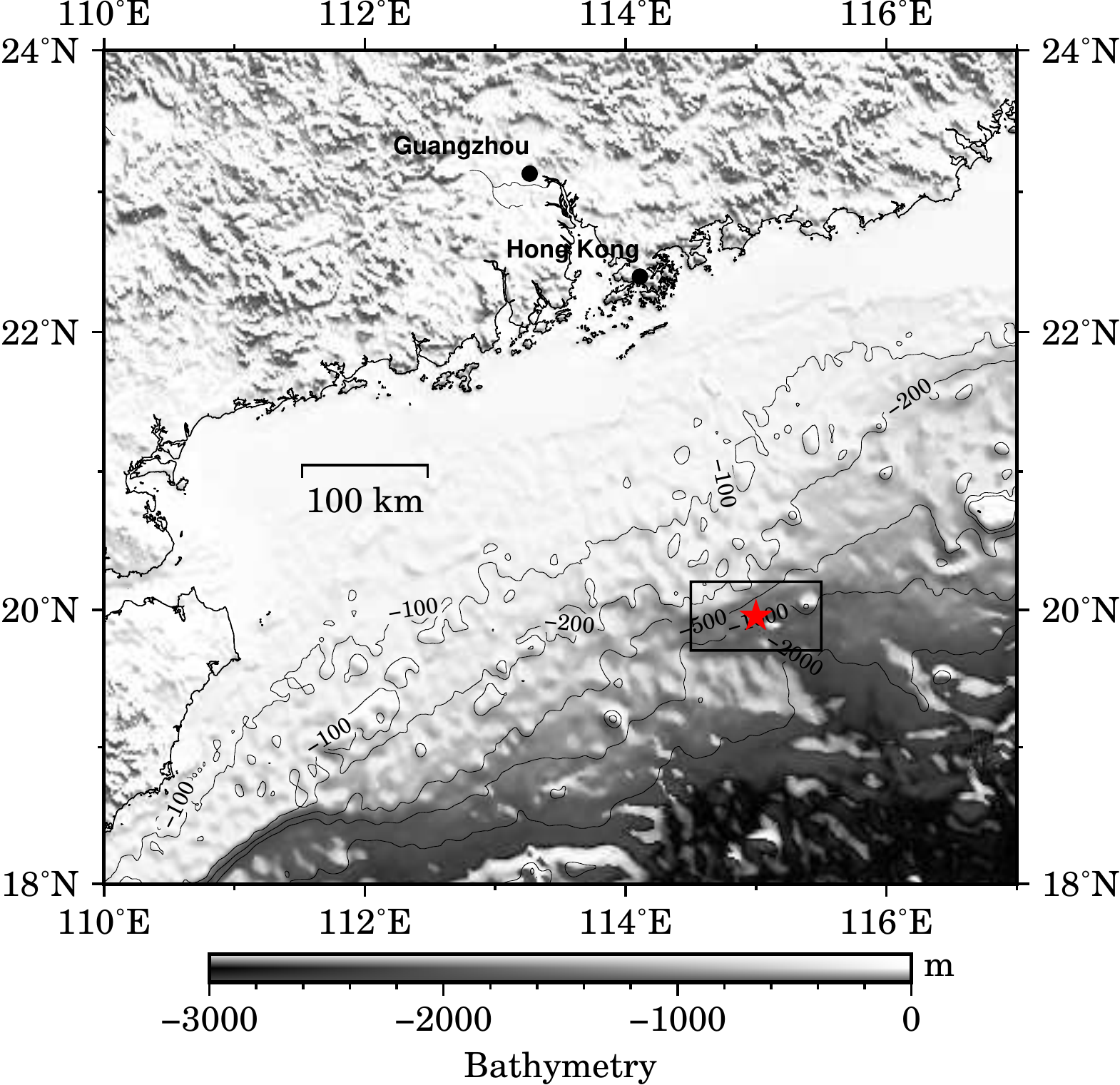}\llap{\parbox[b]{1.5in}{(a)\\\rule{0ex}{2in}}}
 \includegraphics[width=0.45\textwidth]{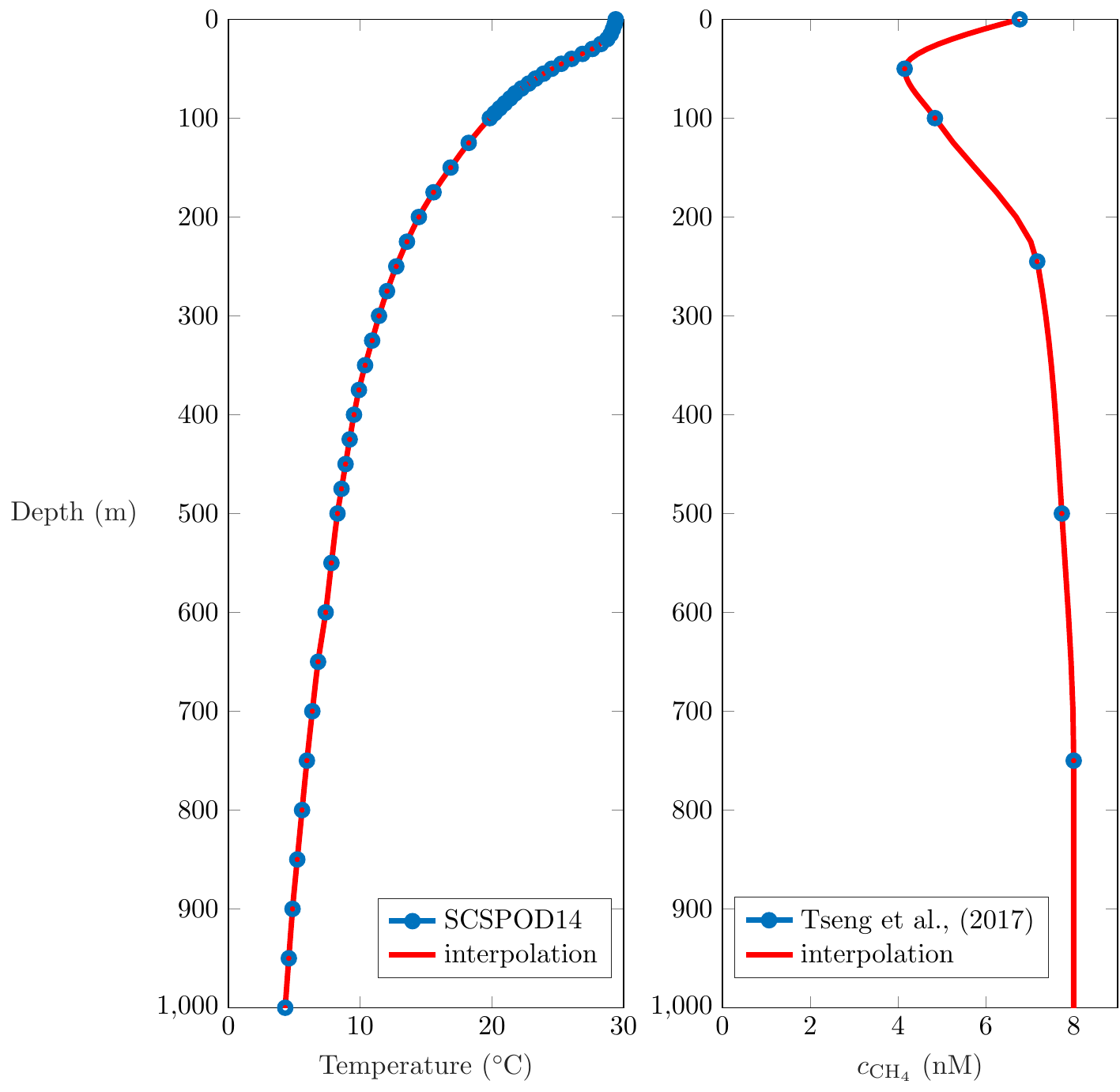}\llap{\parbox[b]{1.9in}{(b)\\\rule{0ex}{2in}}}\llap{\parbox[b]{0.8in}{(c)\\\rule{0ex}{2in}}}
 \caption{(a): the Shenhu area near the Pearl River Mouth Basin, where the star denotes the study site. The temperature of the study site (b) shows that at about \SI{1}{\km} depth, the temperature drops to around \SI{5}{\celsius}, and the dissolved methane concentration (c) plateaus at a high level around \SI{8}{nM}. The temperature and the methane concentration profiles are interpolated with piecewise cubic Hermite interpolating polynomials.}\label{fig:shenhu}
\end{figure}

Taking into account the seasonal change caused by the monsoon, the water temperature and methane concentration near the Shenhu area vary with months. We extract the temperature profile from the SCSPOD14 database \citep{zeng2016scspod14} at a location \ang{19.75} N, \ang{115} E, in the month of July to be consistent with the methane measurement. \Cref{fig:shenhu}(a) shows the study site, and (b) shows the in situ temperature measurement. The methane concentration profile (c) was measured near the study site in July 2005 by \citet{tseng2017methane}. It is notable that the methane concentration is high throughout the depth, approximately \SI{7}{nM} compared with \SIrange{2}{6}{nM} in the Atlantic and Pacific Oceans \citep{reeburgh2007oceanic}, possibly due to methane seeps underneath.

\section{Modeling approach}

The size of the methane bubble is essential to its survival. Decompression during the ascent enlarges the bubble, while the mass transfer to the surrounding seawater reduces the size. For simplicity, the water is treated as stagnant with no upwelling and turbidity currents, and the pressure is hydrostatic. Bubbles are small and sparse enough so the effect of splitting, merging, and turbulence can be neglected. Another assumption is that the bubble remains spherical, and rises vertically. The physical parameters determining the terminal velocity $U$ of a rising methane bubble with a radius $a$ are the densities of the water $\rho$ and the methane gas $\rho'=(1-\beta)\rho$, the viscosity of the water $\mu$, the surface tension $\gamma$, and the gravity $g$. The viscosity of methane gas inside the bubble is negligible compared to the water viscosity, but at large depth the gas density may not be ignored. The transfer of methane across the bubble interface is dependent on the diffusion coefficient $D$ and Henry's law constant $H$. The physical parameters are summarized below in \Cref{table:parameter}. The parameters of water depend on the temperature, but the temperature range in our model is only \SIrange{0}{30}{\celsius}. Within this range, the surface tension and water density change very small and can be treated as constant, while the viscosity can be calculated using the empirical relation \citep{straus1977thermal}
\begin{linenomath*}
 \begin{equation}
  \mu(T)=\mu_0\exp \left(\frac{A}{T-B}\right)
 \end{equation}
\end{linenomath*}
where $\mu_0=\SI{2.414e-5}{\pascal\second}$, $A=\SI{570.58}{\K}$ and $B=\SI{140}{\K}$. The kinematic viscosity of water is $\nu(T)=\mu(T)/\rho$.

\begin{table}[h]
 \centering
 \begin{tabular}{@{}lllll@{}}
  \toprule
  \multicolumn{4}{c}{Model parameters} & Value \\ \midrule
  \multirow{3}{*}{methane gas} & molar mass & $ M $ & [\si{\g/\mole}] & \num{16.04} \\
  & diffusion coefficient & $ D $ & [\si{\m^2/\s}] & \num{1.49e-9} \\
  & Henry's law constant & $ H $ & [\si{\mole\per\pascal\per\cubic\m}] & \num{1.4e-5} \\
  & \multirow{2}{*}{van der Waals constants} & $ A $ & [\si{\bar\liter\squared\per\mole\squared}] & \num{2.303} \\
  & & $ B $ & [\si{\liter\per\mole}] & \num{0.0431} \\ \midrule
  \multirow{3}{*}{water} & density & $ \rho $ & [\si{\kg/\m^3}] & $\sim$\num{1e3} \\
  & dynamic viscosity & $ \mu $ & [\si{\milli\pascal\s}] & \numrange{0.85}{1.8} \\
  & surface tension & $ \gamma $ & [\si{\newton/\meter}] & $\sim0.07$ \\ \bottomrule
 \end{tabular}
 \smallskip
 \caption{Nominal parameter values of the methane gas and seawater. The diffusion coefficient is from \citet{cussler2009diffusion}, the van der Waals constants are from \citet{poling2001properties}, and the Henry's law constant is from \citet{sander2015compilation}.}\label{table:parameter}
\end{table}

\subsection{Dimensional analysis}

From the physical parameters above, some requirements can be obtained if the bubble rises vertically and remains spherical. The main forces at play are the water resistance $F_i\sim\rho U^2d^2$ where $d=2a$, the buoyancy force $F_b\sim g\Delta\rho d^3=g\rho\beta d^3$, the viscous force $F_\mu\sim \mu d U$, and the surface tension force $F_\gamma\sim \gamma d$. From these forces we construct dimensionless parameters including the Reynolds number $\mathrm{Re}=F_i/F_\mu=Ud/\nu$ and the Weber number $\mathrm{We}=F_i/F_\gamma=\rho U^2d/\gamma$. We can also obtain the Morton number $\mathrm{Mo}\equiv g\mu^4/\rho\gamma^{3}\sim\num{1e-10}$, which is the only dimensionless number specific to the water. According to \citet{harper1972motion}, with small \textrm{Mo}, for impure liquids such as the seawater, the marginal instability should occur at $\mathrm{We}<3$ and $\mathrm{Re}\approx200$, which suggests the maximum stable terminal velocity
\begin{linenomath*}
 \begin{equation}
  U_\text{stable}<\frac{\gamma\mathrm{We}}{\mu\mathrm{Re}}\approx\SI{0.58}{\m/\s}.
 \end{equation}
\end{linenomath*}

On the other hand, if the bubble remains spherical, the surface tension force must outweigh the buoyancy force, or the Bond number $\mathrm{Bo}=F_b/F_\gamma\equiv\rho\beta gd^2/\gamma\leq1$, which requires a bubble diameter
\begin{linenomath*}
 \begin{equation}
  d\leq\sqrt{\frac{\gamma}{g\beta\rho}}
 \end{equation}
\end{linenomath*}
while for a rising bubble, the buoyancy force must be greater than the water resistance, or the Froude number $\mathrm{Fr}=F_i/F_b\leq1$,
\begin{linenomath*}
 \begin{equation}
  d\geq\frac{U^2}{g\beta}.
 \end{equation}
\end{linenomath*}
Therefore, if a bubble satisfying the assumptions exists, it is necessary that the two inequalities are compatible, which gives another upper bound of the terminal velocity
\begin{linenomath*}
 \begin{equation}
  U\leq{\left(\frac{\beta g\gamma}{\rho}\right)}^{1/4}.
 \end{equation}
\end{linenomath*}
This velocity, similar to the average drift velocity in two-phase bubbly flows \citep{ishii1979drag} with a difference of a factor of $\sqrt{2}$, is smaller than $U_\text{stable}$ and agrees with the in situ measurements at seep sites \citep{greinert20061300,mcginnis2006fate,sauter2006methane,sahling2009vodyanitskii,leifer2010characteristics,roemer2012quantification,wang2016observations}, so we set the maximum terminal velocity in our model to
\begin{linenomath*}
 \begin{equation}\label{eq:max}
  U_{\max}=\sqrt{2} {\left(\frac{\beta g\gamma}{\rho}\right)}^{1/4}\approx\SI{0.23}{\m/s}.
 \end{equation}
\end{linenomath*}

\subsection{Methane exchange during ascent}

The terminal velocity and mass transfer of a bubble in water has been extensively studied (e.g., \citep{clift2005bubbles}). The correlation of the terminal velocity and the spherical bubble radius can be better expressed by the value of \textrm{Re} and the drag coefficient $C_d$, and the terminal velocity is calculated by
\begin{linenomath*}
 \begin{equation}
  U=\sqrt{\frac{8ga\beta}{3C_d}}.
 \end{equation}
\end{linenomath*}
Numerous experiments have been performed to obtain better correlations \citep[][Table~5.2]{clift2005bubbles} within different ranges of \textrm{Re}, and a unified fit for $\mathrm{Re}<\num{3e5}$ was proposed by \citet{zhang2003kinetics}
\begin{linenomath*}
 \begin{equation}\label{eq:drag}
  C_d=\frac{24}{\mathrm{Re}}\left(1+0.15\mathrm{Re}^{0.687}\right)+\frac{0.42}{1+42500\mathrm{Re}^{-1.16}}
 \end{equation}
\end{linenomath*}
and the equation agrees exceedingly well with the experimental data by \citet{dioguardi2017new}. For stable ascent with a maximum velocity $U_{\max}$, the rising velocity is modified to
\begin{linenomath*}
 \begin{equation}\label{eq:rising}
  U_\text{unified}=\frac{1}{\sqrt{U^{-2}+U^{-2}_{\max}}}=\sqrt{\frac{8ga\beta}{3C_d+2\sqrt{\mathrm{Bo}}}}.
 \end{equation}
\end{linenomath*}

When a single spherical bubble of a radius $a$ rises in the ocean with a velocity $U$, and a diffusion coefficient $D$, the P\'eclet number is calculated by
\begin{linenomath*}
 \begin{equation}
  \mathrm{Pe}=\frac{2aU}{D}
 \end{equation}
\end{linenomath*}
and the mass transfer coefficient $k$ is
\begin{linenomath*}
 \begin{equation}
  k=\frac{\mathrm{Sh} D}{2a}.
 \end{equation}
\end{linenomath*}
where the Sherwood number $\mathrm{Sh}$ depends on the bubble size and velocity. \citet{zhang2003kinetics} proposed a unified equation for $\mathrm{Re}<\num{1e5}$ based on previous piecewise correlations \citep[][Eq. 3-49, 5-25, and Table 5.4]{clift2005bubbles}
\begin{linenomath*}
 \begin{equation}\label{eq:transfer}
  \mathrm{Sh}=1+{(1+\mathrm{Pe})}^{1/3}\left(1+\frac{0.096\mathrm{Re}^{1/3}}{1+7\mathrm{Re}^{-2}}\right)
 \end{equation}
\end{linenomath*}

The amount of methane transferred can be determined by
\begin{linenomath*}
 \begin{equation}
  \dv{n}{t}=4\pi a^2k(c_0-Hp)
 \end{equation}
\end{linenomath*}
where $c_0$ is the concentration of the dissolved methane in the ocean, and the pressure is
\begin{linenomath*}
 \begin{equation}
  p=p_0+\rho g z+\frac{2\gamma}{a}.
 \end{equation}
\end{linenomath*}
The methane in the bubble is treated as van der Waals gas due to the high pressure
\begin{linenomath*}
 \begin{equation}
\left(p+\frac{An^2}{V^2}\right)(V-nB)=nRT
 \end{equation}
\end{linenomath*}
 with the constants $A=\SI{2.303}{\bar\liter\squared\per\mole\squared}$ and $B=\SI{0.0431}{\liter\per\mole}$ \citep{poling2001properties}. The critical pressure for methane is $p_c=A/27B^2=\SI{4.6}{\mega\pascal}$, equivalent to a water pressure $\sim$\SI{460}{mbsf}. The critical temperature is $T_c=8A/27RB=\SI{190}{\K}$, so the methane remains gaseous, and the gas density in the bubble is related to the molar mass $M$ by $\rho'=nM/V$. The bubble vertical motion can be described by
\begin{linenomath*}
 \begin{equation}
  \dv{z}{t}=-U.
 \end{equation}
\end{linenomath*}

We numerically solve the equations above, and calculate the amount of methane gas remained for those bubbles that make to the surface.

\section{Results and discussion}

\subsection{Rising of the methane bubble}

\begin{figure}[h]
 \includegraphics[width=0.9\textwidth]{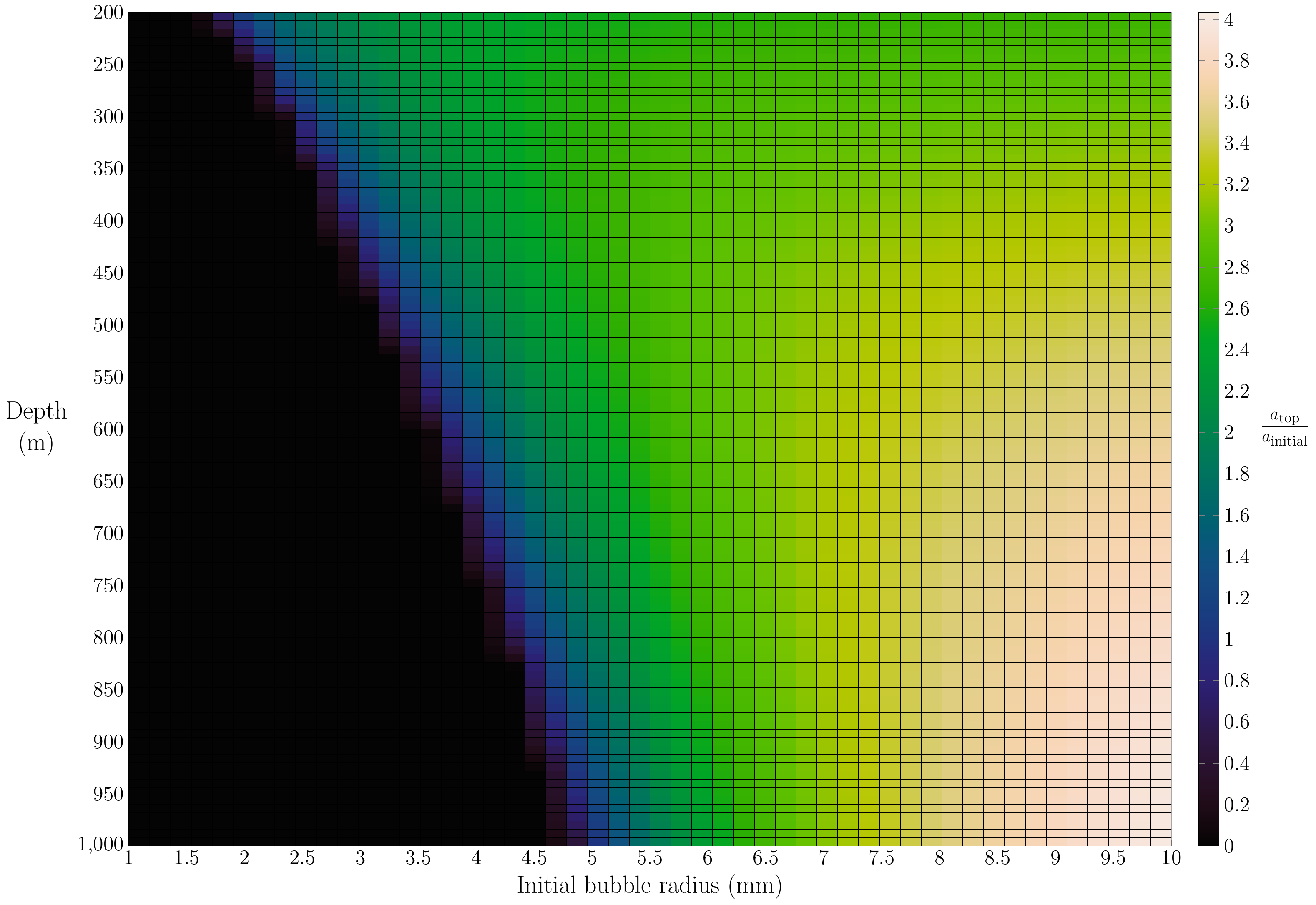}\llap{\parbox[b]{5in}{(a)\\\rule{0ex}{3in}}}
 \includegraphics[width=0.9\textwidth]{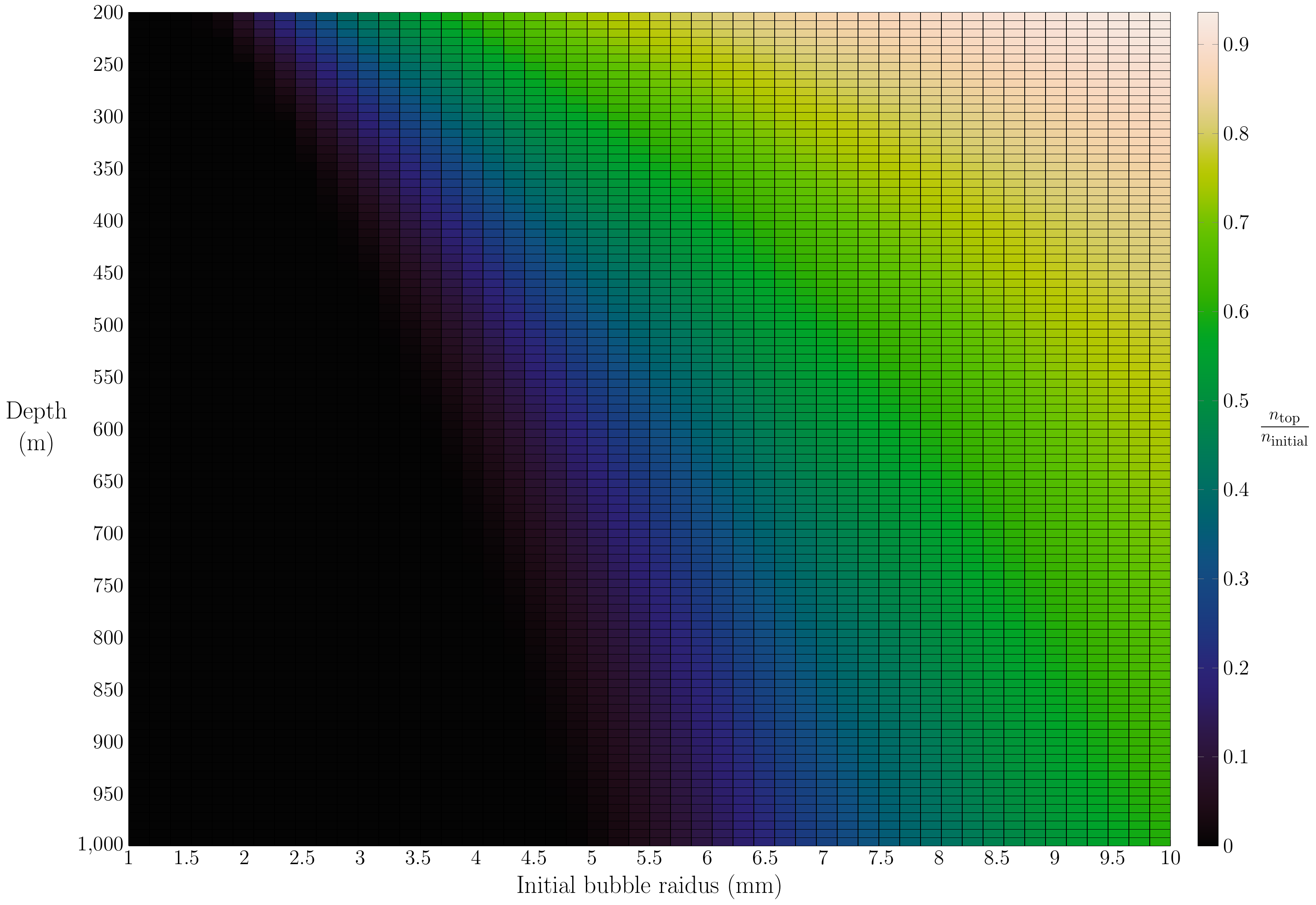}\llap{\parbox[b]{5in}{(b)\\\rule{0ex}{3in}}}
 \caption{(a) The ratio of the radius of the final bubble reaching the atmosphere to the initial bubble radius. (b) The ratio of the remaining methane gas reaching the atmosphere. In shallow water in the Shenhu area, a large bubble can expand to almost four times in the radius, and bring more than \SI{90}{\percent} of the initial methane gas to the atmosphere.}\label{fig:bubble}
\end{figure}

We simulated the ascent of bubbles released at depths between \SIrange{200}{1000}{\m}, with a initial radius between \SIrange{1}{10}{\mm} using the temperature and methane concentration for the Shenhu area. \Cref{fig:bubble}a shows the ratio of the final bubble radius to the initial bubble radius $a_\text{top}/a_\text{initial}$, and \Cref{fig:bubble}b shows the ratio of remaining methane $n_\text{top}/n_\text{initial}$ of the methane bubbles. Very small bubbles are not likely to reach the surface. For example, a bubble with a radius of \SI{2}{\mm} rising from \SI{200}{\m} depth will lose about \SI{90}{\percent} of the methane when it reaches the atmosphere. Large bubbles, however, will retain more gas, and the expansion is significant due to the decompression. A bubbles with a initial radius over \SI{8}{\mm} rising from depths between \SIrange{450}{700}{\m} can expand as many as four times in the radius when it makes to the surface.

\subsection{Instability of bubble rising path}

The velocity of a large bubble increases as it expands, and the Reynolds number \textrm{Re} can grow large enough that the assumptions of stable ascent become invalid, and the rising path no longer maintains stable. A bubble with a radius of \SI{9}{\mm} rising at the terminal speed of $U\approx\SI{0.16}{\m/\s}$ in the water of \SI{27}{\celsius} corresponds to $\mathrm{Re}\sim\num{3.3e3}$, and $\mathrm{We}\approx3.2$, greater than the criteria of marginal stability $\mathrm{We}<3$ and $\mathrm{Re}\approx200$ \citep{harper1972motion}. Because the bubble is unlikely to grow beyond \SI{5}{\cm} in the radius in our simulation, which corresponds to a Galilei number $\mathrm{Ga}\equiv \sqrt{ga}a/\nu\sim40$ as the transition to oscillatory ellipsoidal or spherical caps according to \citet{tripathi2015dynamics}, the bubbles are still spherical, but they may split into smaller bubbles.

\subsection{Formation of methane hydrate shells}

\begin{figure}[h]
 \centering
 \includegraphics[width=0.8\textwidth]{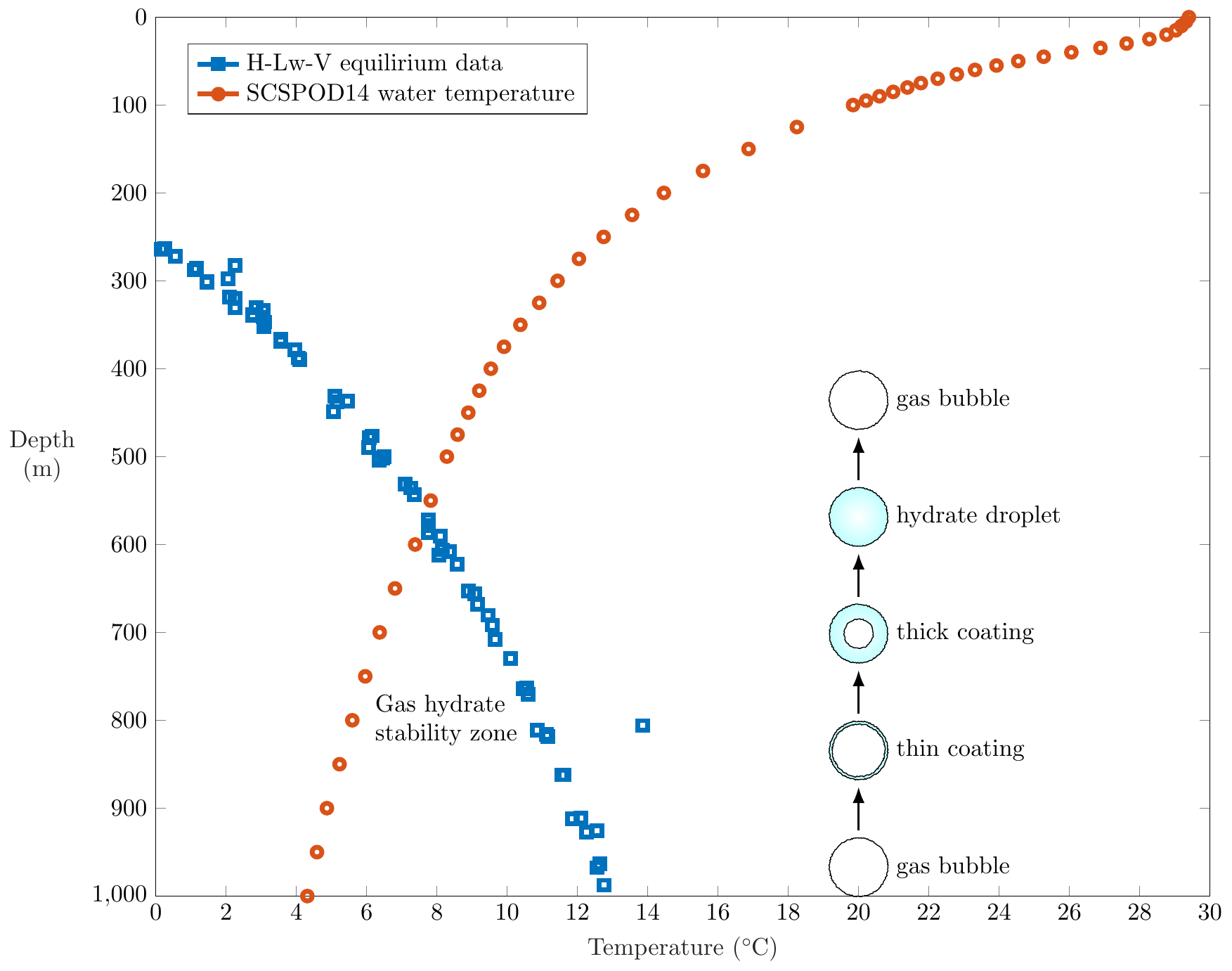}
 \caption{The gas hydrate stability zone in the Shenhu area bounded by the phase equilibrium curve and the SCSPOD14 water temperature profile. The top of GHSZ in our study site is at about \SI{600}{\m}. A schematic shows the formation and dissociation of the hydrate shell around the methane bubble when it rises through the GHSZ.}\label{fig:shell}
\end{figure}

Within the gas hydrate stability zone (GHSZ), a coating of hydrate shell encapsulates the methane bubble \citep{maini1981experimental,rehder2002enhanced}, and exists until top of the stability zone is reached \citep{rehder2002enhanced,warzinski2014dynamic}. The top limit of the GHSZ can be determined by the intersection of the methane phase equilibrium curve \citep[see data compiled by][]{sloan2007clathrate} and the temperature profile (\Cref{fig:shell}). When the hydrate shell forms, water filters through the porous shell, and increases the total mass of the bubble-converted droplet \citep{ribeiro2008modelling}. The shell also provides additional pressure to the gas within, so the droplet does not expand as a bubble should. As a result, the hydrate shell will help to keep more methane in the bubble.

\subsection{Non-ideal gas}

The methane gas is treated as van der Waals gas in our simulation, but the equation of state of methane under large pressure is much more complicated. It is possible to use a more sophisticated equation of state such as the model proposed by \citet{peng1976new}, but since the temperature is around $\SI{0}{\celsius}\gg T_c=\SI{190}{\K}$, the deviation is very small. Also, we assume constant diffusion coefficient, surface tension and Henry's law constant, but it is possible that these parameters also change as the methane becomes non-ideal.

\subsection{Turbidity current and upwelling flux}

We treated the seawater as stagnant with no motion, however, the water movement is nowhere near stillness due to upwelling fluxes from underneath and turbidity currents from the continental slope. These flows will disturb the water body, and enhanced mixture from the flows will facilitate the dissolution of the methane gas, and reduce the amount of methane reaching the atmosphere.

\section{Conclusion}

We simulated the ascent of methane bubbles in the Shenhu area, and calculated the amount of methane that reaches the atmosphere. Our simulation shows that for the Shenhu area, methane gas emitted from deep sites $>$\SI{1}{\km} is not likely to reach the ocean surface, as indicated in \Cref{fig:bubble}, as long as the bubble radius is smaller than \SI{5}{\mm}. Hydrate-coated bubbles can survive longer and retain more more methane, but still small bubbles cannot make through the water column. Kilometers of water column is very effective in keeping the methane in the water body, and a gas blowout will not result in an abrupt amount of methane entering the atmosphere. However, elevated methane emission will lead to larger bubbles, and as the excessive methane is dissolved in shallower water, subsequent anaerobic methane oxidation will change the distribution of pH, which has negative impact on the ocean ecology. Small bubbles with radii \SIrange{2}{3}{\mm} released between \SIrange{200}{1000}{\m} will entirely dissolve in the ocean, and large bubbles over \SI{5}{\mm} rising from \SI{500}{\m} depth can keep at least \SI{40}{\percent} when it makes to the ocean surface. In our study area, the top of gas hydrate stability zone lies at about \SI{600}{\m}, so a methane bubble released below this depth can survive longer distance and transfer more methane to shallower depths.

\acknowledgments{}
= enter acknowledgments here =

\bibliography{ref}

\listofchanges{}

\end{document}